# On the relation between reliable computation time, float-point precision and the Lyapunov exponent in chaotic systems

Wang PengFei[1,2] and Li JianPing[3]

1 Center for Monsoon System Research, Institute of Atmospheric Physics, Chinese Academy of Sciences, Beijing 100190, China

2 State Key Laboratory of Numerical Modeling for Atmospheric Sciences and Geophysical Fluid Dynamics, Institute of Atmospheric Physics, Chinese Academy of Sciences, Beijing 100029, China

3 College of Global Change and Earth System Science, Beijing Normal University, 100875, China

Corresponding author: wpf@mail.iap.ac.cn

**Abstract** The relation among reliable computation time, $T_c$, float-point precision, $K$, and the Lyapunov exponent, $\lambda$, is obtained as $T_c= (\ln B/\lambda)K+C$, where $B$ is the base of the float-point system and $C$ is a constant dependent only on the chaotic equation. The equation shows good agreement with numerical experimental results, especially the scale factors.

## 1. Introduction

Obtaining the long-term true trajectories of chaotic dynamical systems by numerical approaches is a complex task. Researchers have reported [1-3] that numerical methods can provide approximate trajectories close to reality by applying Riemannian manifolds theory [4]. However, a remaining problem is how well and how long the numerical trajectory approximates the real one [5]. Many studies have documented the sensitivities of computation parameters for numerical solutions of chaotic equations [6-11], and these studies indicate that, in spite of there being no initial errors, the computation is still limited by the maximal effective computation time ($T_c$) due to round-off error. Li et al. [7] carried out systematic investigations on these phenomena for nonlinear ordinary differential equations (ODEs), by employing numerical experiments and an analytical approach. The sensitivity of the computed results for chaotic systems is important. For example, previous results [6, 9, 10] have indicated that the maximal effective computation time is approximately 35 LTU (Lorenz time unit) for Lorenz equations under double precision. Nevertheless, studies are still needed to carry out theoretical analyses of data for computation times of longer than 35 LTU. It is important to design difference solvers to provide the correct solution for times beyond 35 LTU, which is the situation for many types of chaotic systems.

The existence of $T_c$ indicates that if we need the solution of $t = 1000$, single- and double-precision computers are insufficient. In order to overcome the shortcomings of computation precision, a suite of multiple-precision (MP) software [12] has been developed. This library can provide user-defined floating-point precision in the computation. Using MP, it is possible to choose a sufficiently high precision level with a certain step size, $h$, to maintain round-off errors that are negligible compared with the truncation error. Wang et al. [13] and Liao [10] have shown these MP libraries to solve Lorenz equations, demonstrating that high precision is valuable in terms of obtaining correct numerical solutions. However, a problem while applying MP is that when the precision increases $n$ times, the speed of computation will decrease $n$ times. In addition, unnecessarily high precision (beyond the scope of the problem) will waste a large amount of computer memory. Therefore, choosing a suitable level of precision for the computation is a challenging task.

To date, the available methods for estimating the computation precision generally depend on numerical experiments. For example, Liao [10] obtained a relation from reliable numerical experiments as

$$T_c \approx 2.5K - 4.26, \qquad (1)$$

where $T_c$ denotes the reliable computation time and $K$ is the significant digital number of the float-point computation base 10. In fact, relation (1) is a special case of that obtained by Li et al. [7], but the formula they obtained needs to determine some parameter values, and thus is not as convenient as formula (1) in practice.

In this paper, we apply statistical and error propagation theory to obtain a more useful relation between the computation time, precision, and Lyapunov exponent for the numerical integration of chaotic systems.

## 2. Theoretical analysis

Here, we demonstrate the precision estimated by analyzing Lorenz's [25] equation:

$$\begin{cases} \dfrac{dx}{dt} = -\sigma x + \sigma y \\ \dfrac{dy}{dt} = Rx - y - xz, \\ \dfrac{dz}{dt} = xy - bz \end{cases} \qquad (2)$$

where $R$, $\sigma$, and $b$ ($R = 28.0$, $\sigma = 10.0$, $b = 8/3$) are constants, and $t$ is nondimensional time.

Let $\lambda$ the characterize Lyapunov exponent (or maximal Lyapunov exponent) for the chaotic dynamical system (2), and $\lambda > 0$. When we integrate this chaotic system and only consider the propagation of round-off error (high enough order methods, such as the Taylor series method, can guarantee the truncation error negligible compare to round-off error), the initial value is $x_0$, and in the first step, the theoretical value is $x_1$. The numerical value can be expressed as

$$x_1(1 + c_1\varepsilon), \qquad (3)$$

where $\varepsilon = 10^{-K}$ denotes the relative error, $K$ is the significant digital number, $-\dfrac{1}{2} \leq c_i \leq \dfrac{1}{2}$ is a parameter representing the random of round-off error, and it satisfies the uniform distribution. The error between the numerical solution and the theoretical value is $x_1 c_1 \varepsilon$. In the second step, when computing the solution $x_2$, it can be regarded as a nonlinear error propagation problem with initial error $\delta_0 = x_1 c_1 \varepsilon$. Since the round-off error possesses randomness, the numerical solution is $x_2' = \left(x_2 + x_1 c_1 \varepsilon e^{\lambda h}\right)(1 + c_2 \varepsilon)$. The error in the second step is $x_2' - x_2 = x_1 c_1 \varepsilon e^{\lambda h} + x_2 c_2 \varepsilon + c_1 c_2 \varepsilon^2 e^{\lambda h}$. The $\varepsilon$ is a small-scale value, and thus we can omit the second order part, $c_1 c_2 \varepsilon^2 e^{\lambda h}$, and obtain

$$x_2' - x_2 = x_1 c_1 \varepsilon e^{\lambda h} + x_2 c_2 \varepsilon. \qquad (4)$$

In the third step, we can compute the value $x_3$ from an initial error

$\delta_0 = x_1 c_1 \varepsilon e^{\lambda h} + x_2 c_2 \varepsilon$, and by repeating this procedure we obtain the *n*-th step's error formula:

$$E_n \equiv x_n' - x_n = x_1 c_1 \varepsilon e^{\lambda h(n-1)} + x_2 c_2 \varepsilon e^{\lambda h(n-2)} + \cdots + x_n c_n \varepsilon. \tag{5}$$

The average value of $E_n$ is 0, but when we consider the variation of $E_n$ we find that

$$|E_n|^2 = \left(x_1 c_1 \varepsilon e^{\lambda h(n-1)} + x_2 c_2 \varepsilon e^{\lambda h(n-1)} + \cdots + x_n c_n \varepsilon\right)^2$$
$$= \left(x_1 c_1 \varepsilon e^{\lambda h(n-1)}\right)^2 + \left(x_2 c_2 \varepsilon e^{\lambda h(n-1)}\right)^2 + \cdots + \left(x_n c_n \varepsilon\right)^2 + 2 \sum_{\substack{0 \le i,j \le n-1 \\ i \ne j}} \left(x_i c_i \varepsilon e^{\lambda h(n-i)}\right)\left(x_j c_j \varepsilon e^{\lambda h(n-j)}\right).$$

In order to evaluate the expected value of $|E_n|^2$, we first integrate it in the $c_i$ space,

$$\overline{|E_n|^2} = \int_{-\frac{1}{2}}^{\frac{1}{2}} \int_{-\frac{1}{2}}^{\frac{1}{2}} \left(x_1 c_1 \varepsilon e^{\lambda h(n-1)}\right)^2 + \left(x_2 c_2 \varepsilon e^{\lambda h(n-1)}\right)^2 + \cdots + \left(x_n c_n \varepsilon\right)^2 + 2 \sum_{\substack{0 \le i,j \le n-1 \\ i \ne j}} \left(x_i c_i \varepsilon e^{\lambda h(n-i)}\right)\left(x_j c_j \varepsilon e^{\lambda h(n-j)}\right) dc_i dc_j$$
$$= \frac{1}{12} \sum_{i=1}^{n} \left(x_i \varepsilon e^{\lambda h(n-i)}\right)^2 + 2 \sum_{\substack{0 \le i,j \le n-1 \\ i \ne j}} \left(x_i \varepsilon e^{\lambda h(n-i)}\right)\left(x_j \varepsilon e^{\lambda h(n-j)}\right),$$

and then integrate it in the $x_i$ space. We note that $\int_{x_a}^{x_b} p(x) x \, dx = 0$ (e.g., in a Lorenz system, the variables *x* and *y* are symmetrical about 0, but for variable $z$ $\int_{z_a}^{z_b} p(z) z \, dz \ne 0$, we should in this situation use the formula in the appendix), where $x_a, x_b$ denote the lower and upper bounds of variable $x$, and $p(x)$ is the statistical distribution function corresponding to *x*.

$$\overline{|E_n|^2} = \int_{x_a}^{x_b} \int_{x_a}^{x_b} \left\{ \frac{1}{12} \sum_{i=1}^{n} \left(x_i \varepsilon e^{\lambda h(n-i)}\right)^2 + 2 \sum_{\substack{0 \le i,j \le n-1 \\ i \ne j}} \left(x_i \varepsilon e^{\lambda h(n-i)}\right)\left(x_j \varepsilon e^{\lambda h(n-j)}\right) \right\} p(x_i) p(x_j) dx_i dx_j$$
$$= \frac{1}{12} \left\{ \left(\varepsilon e^{\lambda h(n-1)}\right)^2 + \left(\varepsilon e^{\lambda h(n-1)}\right)^2 + \cdots + \left(\varepsilon\right)^2 \right\} \int_{x_a}^{x_b} p(x) x^2 dx$$

which can be concluded as

$$\overline{|E_n|^2} = \frac{1}{12} \overline{x^2} \varepsilon^2 \frac{e^{2\lambda h n} - 1}{e^{2\lambda h} - 1}. \tag{6}$$

We choose the norm of $|E_n|^2$ and set the error tolerance to $\delta$. Thus,

$$\delta = \sqrt{\frac{1}{12} \overline{x^2}} \varepsilon \sqrt{\frac{e^{2\lambda h n} - 1}{e^{2\lambda h} - 1}}.$$

Since $hn = T_c$, $e^{2\lambda hn} - 1 \approx e^{2\lambda hn}$, such that

$$\delta = \varepsilon \sqrt{\frac{1}{12}\overline{x^2}} \frac{e^{\lambda T_c}}{\sqrt{e^{2\lambda h} - 1}}.$$

We perform the logarithm in this equation and obtain

$$T_c = \frac{1}{\lambda} \ln \frac{\sqrt{e^{2\lambda h} - 1}\delta}{\sqrt{\frac{1}{12}\overline{x^2}}} - \frac{1}{\lambda} \ln \varepsilon.$$

Now, we take $\varepsilon = 10^{-K}$ into the formula and obtain the relation as

$$T_c = \frac{\ln 10}{\lambda} K + \frac{1}{\lambda} \ln \frac{\sqrt{e^{2\lambda h} - 1}\delta}{\sqrt{\frac{1}{12}\overline{x^2}}}. \tag{7}$$

Denote

$$C = \frac{1}{\lambda} \ln \frac{\sqrt{e^{2\lambda h} - 1}\delta}{\sqrt{\frac{1}{12}\overline{x^2}}}, \tag{8}$$

and the more general relation formula can be expressed as

$$T_c \approx \frac{\ln B}{\lambda} K + C, \tag{9}$$

where $B$ is the base of the float-point system.

We can evaluate the accuracy of formula (9). For example, in a Lorenz system that has $\overline{x^2} \approx 62.87$, $\lambda = 0.906$, if we choose $h = 0.01, \delta = 1$, the computed value $C \approx -3.12$, and $\frac{\ln B}{\lambda} = \frac{\ln 10}{0.906} \approx 2.51$. These two values match well with Liao's [10] formula (1), especially the value of $\frac{\ln B}{\lambda}$, which is almost the same as the experiment's value.

The parameter $C$ depends partly on the step size, $h$, and the error tolerance, $\delta$. Thus, different choices will cause different values. Even though formula (9) has the main part $\frac{\ln B}{\lambda} K$, especially when $K$ is a large integer, in this situation the evaluation error of parameter $C$ has a tiny effect on $Tc$. Furthermore, the parameter $C$ has connections with the object variable, which is not mentioned in Liao's [10] and Li's [7] previous studies. For example, if we study the variable *z* of a Lorenz system, the formula of *C* will be more complicated due to the expected value of *z* being non-zero.

## 3. Conclusion and discussion

We obtain the value of the scale factor of *K* for some famous chaotic systems [14-19] using the parallel multiple-precision Taylor scheme [20, 21].

**Table 1.** The $T_c - K$ formula by experiments and the theoretical value of $\frac{\ln B}{\lambda}$, where *K* denotes the base 10 precision and $K_b$ the base 2 precision.

| Chaotic system | $T_c - K$ formula by experiments | h | $\lambda$ | $\frac{\ln B}{\lambda}$ |
|---|---|---|---|---|
| Lorenz | $T_c = 2.50K - 4.26$ | 0.01 | 0.906 | 2.51 |
| Chen | $T_c = 0.33K_b - 2.8$ | 0.01 | 2.027 | 0.34 |
| Rossler | $T_c = 9.10K_b - 60$ | 0.01 | 0.076 | 9.12 |
| Coupled Lorenz | $T_c = 0.06K_b - 0.3$ | 0.001 | 11.50 | 0.06 |
| Lü | $T_c = 0.50K_b$ | 0.01 | 1.380 | 0.50 |

From Table 1, we can conclude that the value $\frac{\ln B}{\lambda}$ matches well with the scale factor values in experiments of the $T_c - K$ formula.

Formula (8), used to express *C*, still possesses some error. In the Lorenz system, the value of *C* has about a 25% difference with numerical experiments. We suggest the following possible reasons for this difference:

1. When *M* is not large, the value of $T_c$ is also not large, and then the value of $\overline{x^2}$ has some local property because it is not ergodic in the whole attractor.

2. We do not consider the power 2 and upper items, such as $\varepsilon^2$, in formula (4).

3. We assume the statistical distribution function for round-off error is uniform, but some studies indicate it sometimes follows a more complex distribution function.

4. The approximate round-off error formula in (3) only considers the effect of *x*, but for computation schemes such as the Taylor and Runge-Kutta scheme, the round-off part should be proportional to the function of $f(x, y, z)$ in a three-variable ODE system.

Formula (7) has another use; that is, we can estimate the maximal Lyapunov exponents for some complex ODE by numerical experiments to fit the correct $T_c \sim K$ diagram.

**Appendix**

For the case $\int_{x_a}^{x_b} xp(x)dx \neq 0$, we can obtain the round-off error formula as

$$\overline{|E_n|^2} = \frac{1}{12}\overline{x^2}\varepsilon^2 \frac{e^{2\lambda hn}-1}{e^{2\lambda h}-1} + \overline{x}^2\varepsilon^2 \frac{2(e^{\lambda hn}-1)^2}{(e^{2\lambda h}-1)(e^{\lambda h}-1)}.$$

Since $hn = t$, $e^{2\lambda hn} - 1 \approx e^{2\lambda t}$,

$$\overline{|E_n|^2} = \frac{1}{12}\overline{x^2}\varepsilon^2 \frac{e^{2\lambda t}}{e^{2\lambda h}-1} + \overline{x}^2\varepsilon^2 \frac{2e^{2\lambda t}}{(e^{2\lambda h}-1)(e^{\lambda h}-1)}.$$

**Acknowledgements**: This research was jointly supported by the National Natural Sciences Foundation of China (41375112) and the National Basic Research Program of China (2011CB309704).